\documentclass{elsart}
\usepackage{natbib}
\usepackage{epsfig}
\input{epsf.sty}
\usepackage{epsfig}
\usepackage{longtable}

\begin{document}
\runauthor{Xu}
\begin{frontmatter}

\title{A polytropic model of quark stars}

\author{X. Y. Lai and R. X. Xu}\footnote{
Corresponding author.\\
{\em Email address:} r.x.xu@pku.edu.cn.}

\address{School of Physics and State Key Laboratory of Nuclear
Physics and Technology, Peking University, Beijing 100871, China}

\begin{abstract}
A polytropic quark star model is suggested in order to establish a
general framework in which theoretical quark star models could be
tested by observations.
The key difference between polytropic quark stars and the polytropic
model studied previously for normal (i.e., non-quarkian) stars is
related to two issues: (i) a constant term representing the
contribution of vacuum energy may be added in the energy density and
the pressure for a quark star, but not for a normal star; (ii) the
quark star models with non-vanishing density at the stellar surface
are not avoidable due to the strong interaction between quarks.
The first one implies that the vacuum inside a quark star is
different from that outside, while the second one is relevant to the
effect of color confinement.
The polytropic equations of state are stiffer than that derived in
conventional realistic models (e.g., the bag model) for quark
matter, and pulsar-like stars calculated with a polytropic equation
of state could then have high maximum masses ($>2M_\odot$).
Quark stars can also be very low massive, and be still
gravitationally stable even if the polytropic index, $n$, is greater
than 3.
All these would result in different mass-radius relations, which
could be tested by observations.
In addition, substantial strain energy would develop in a solid
quark star during its accretion/spindown phase, and could be high
enough to take a star-quake.
The energy released during star-quakes could be as high as $\sim
10^{47}$ ergs if the tangential pressure is $\sim 10^{-6}$ higher
than the radial one.

\vspace{5mm} \noindent {\it PACS:}
97.60.Gb, 97.60.Jd, 95.30.Cq%
\end{abstract}


\end{frontmatter}


\section{Introduction}

It depends on the state of matter at supra-nuclear density to model
pulsar's structure, which is unfortunately not certain due to the
difficulties in physics although some efforts have been made for
understanding the behavior of quantum chromo-dynamics (QCD) at high
density.
Of particular interest is whether the density in such compact stars
could be high enough to result in unconfined quarks (quark matter).
Stars composed of quarks (and possible gluons) as the dominant
degrees of freedom are called quark stars, and there is possible
observational evidence that pulsar-like stars could be quark
stars~\citep[see reviews, e.g.,][]{weber05,xu08a,xu08b}.
But it is still a problem to model a realistic quark star for our
lack of knowledge about the real state of quark matter.

The study of cold quark matter opens a unique window to connect
three active fields: particle physics, condensed matter physics, and
astrophysics.
Many possible states~\citep[see, e.g.,][]{csc08} of cold quark
matter are proposed in effective QCD models as well as in
phenomenological models.
An interesting suggestion is that quark matter could be in a solid
state~\citep{xu03,Horvath05,Owen05,mrs07}. Solid relativistic stars
are challenging astrophysicists since the stelar matter can not be
well approximately by a perfect fluid and the conventional TOV
(Tolman-Oppenheimer-Volkov) equation is thus not applicable.
Nevertheless, in case of static and spherically symmetric gravity,
the equilibrium equation could be similar to the TOV equation, by
introducing a deviation between radial and tangential
pressures~\citep[see, e.g.,][]{xty06}.
However, one has also to know the radial pressure, $P$, as a
function of density $\rho$ (and possible other parameters) in order
to model a quark star in a solid state.

No realistic relation of $P(\rho)$ is available since no cold quark
matter has been discovered experimentally and/or observationally
with certainty, although many modeled relations between $P$ and
$\rho$ are proposed in the literatures. Among the relations, a class
of linear equations of state, $P=\kappa(\rho-\rho^{\prime})$, is
currently focused, with two free parameters $\kappa$ and
$\rho^{\prime}$~\citep[see, e.g.,][]{zdunik00,sm07},
in the framework of the bag model.
Both relations derived in the bag model and in the density-dependent
quark model can be regarded as special cases of the linear relation
of $P(\rho)$.
Whatsoever, the linear equation could not be adequate if possible
quark-clustering occurs in cold quark matter~\citep{xu03}. Such
matter with clustered quarks could be in a fluid state at high
temperature but in a solid state at sufficient low temperature.
It should be worth noting that the interaction between quarks in a
fireball with quarks and gluons is still very strong~\citep[i.e.,
the strongly coupled quark-gluon plasma,][]{Shuryak}, according to
recent achievements of relativistic heavy ion collision experiments.
Such a strong coupling may naturally render quarks grouped in
clusters.
How can then one state a reliable $P$-$\rho$ relation in order to
establish a framework in which theoretical stellar models could be
tested by observations if the quark-clustering effect is included?

Astronomers faced a similar problem when trying to model
non-quarkian normal stars
(e.g., main sequent stars and white dwarfs).
A polytropic model with equation of state, $P=K\rho^\Gamma$
($\Gamma=1+1/n$), had been extensively studied previously~\cite[see,
e.g.,][]{Chandrasekhar} for main sequent stars as well as white
dwarfs under the Newtonian gravity.
This model has also been extended under general
relativity~\citep{Tooper,Shapiro Teukolsky}. The polytropic models
are valuable because they could help us to model stars composed of
realistic matter, such as ideal gas, photon gas, and degenerate
fermi gas.
Would it be possible for us to do a parallel investigation for quark
stars?
As discussed previously, QCD, which is still developing in
low-energy regime, should be involved to describe cold quark matter
and the equation of state of quark stars is then uncertain
up-to-now.
Furthermore, it might be problematic to calculate the state of
macroscopic quark stars in QCD because QCD is till a local
theory~\citep{Qiu}.
Nonetheless, if quarks are clustered in quark stars where quarks are
coupled strongly, the state of cold quark matter might be
approximated phenomenologically by polytropic equations of state,
since one may draw naively an analogy between the clusters in quark
matter and the nuclei in normal matter.

Some authors had applied the polytropic equations of state to model
hybrid stars which have cores composed of unconfined
quarks~\citep{lin06,zdunik06}.
However, their polytropic equations of state are only used to
describe normal phase (composed of normal baryons) and mixed phase
(composed of denser baryon matter); for the quark phase inside the
core, the equation of state is still the form of the bag model.
In this paper, we will apply for the first time the polytropic
equations of state for quark matter and calculate the structures of
quark stars, with different polytropic indices, $n$.
These stellar models could also be regarded as an extension to the
quark star models with linear equations of state.

We are going to model quark stars in two separated ways. (i) The
vacuum energy inside and outside quark matter is different. As
proposed by~\cite{Bambi}, quark stars could represent an laboratory
to investigate the cosmological constant problem if some compensate
field exist to generate a constant-like term to compensate the
difference of vacuum energy, and the mass-radius curves of quark
stars will be different from the standard case.
We generalize Bambi's idea to discuss generic polytropic equations
of state.
(ii) The vacuum inside and outside of a quark star is assumed the
same, i.e., quark stars have no QCD vacuum energy. In both cases, a
key difference between polytropic quark star and normal star models
lies on the surface density $\rho_{\rm sur}$ ($\rho_{\rm sur}
> 0$ for the former but $\rho_{\rm sur} = 0$ for the latter), since a quark
star could be bound not only by gravity but also by additional
strong interaction due to the strong confinement between quarks.
Analogously, a quark stars without QCD vacuum energy could be
similar to an asteroid with a sharp surface where the density is
also none-zero.
The non-zero surface density is still natural in the case with the
linear equation of state, where the binding effect is represented by
the bag constant, $B$ (and then $\rho_{\rm sur}=4B$).

The stability of a polytropic star depends also on
the surface density. It is well known that a normal star with zero
surface density should be unstable if $n>3$ in the Newtonian
gravity, and that the case of $n=3$ is still unstable in general
relativity~\cite[see details in, e.g.,][]{Shapiro Teukolsky}.
However, in the models we will demonstrate, a quark star could still
be stable even if $n>3$.

The structures of related compact stars have been studied in general
relativity by some authors.
\cite{db1983} derived an analytical expression of mass-radius
relation for isotropic stars in general relativity, and Herrera et
al.~\citep{hs1997,hb2004} analyzed a set of solutions to the
Einstein's equations for anisotropic matter.
It is very interesting to model stars with anisotropic pressure for
some physical reasons~\citep{Ruderman,Sawyer,Sokolov}.
Harko and Mak~\citep{Harko 2,Harko 1} derived an analytical
expression of mass-radius relation for anisotropic stars in general
relativity, and discussed the constraints for the anisotropic
parameter.
They had also presented an exact analytical solution of the
gravitational equations describing a static spherically symmetric
anisotropic quark matter distribution~\citep{Harko 3}.
These authors did not start with an equation of state, but studied
the density and pressure in a more general framework based on the
energy-momentum tensor.
It is worth noting that the results obtained by above authors are
still parameter- and assumption-dependent, even for the so-called
exact solutions.
We will alternatively study the problem, with an explicit form of
equation of state.
We also consider the case of anisotropic stars and compute the
gravitational energy released during quakes of solid quark stars,
with a parameter range given by~\cite{Harko 2}.
Numerical results show that, if the tangential pressure which is
slightly larger than the radial one changes abruptly, the
gravitational energy released could be high enough to power the
supergiant flares observed from soft $\gamma$-ray repeaters.

This paper is arranged as follows. The details of polytropic model
of quark stars of perfect and unperfect fluids, respectively, are
presented in \S2 and \S3. The numerical results are shown in \S4.
The paper is concluded in \S5.

\section{Stars of Perfect fluid}

\subsection{Quark Stars without QCD Vacuum Energy}

If there is no difference between the vacuum inside and outside of a
quark star, the equation of state for a quark star is the standard
polytropic model, with a non-zero surface density, representing the
strong confinement between quarks.
In this point of view, quark stars could analogously be similar to
asteroids: the electromagnetic force dominates over gravity in the
later, while the strong interaction can not be negligible in the
former. Consequently, both of those objects can have a sharp surface
where the density goes down to zero in a negligible small scale.

Stars of perfect fluid in general relativity were discussed
by~\cite{Tooper}, with an equation of state,
\begin{equation} P=K{\rho_g}^\Gamma, \end{equation} \begin{equation} \rho=\rho_g c^2+n P,
\end{equation}
where $\rho_g$ is the part of the mass density which satisfies a
continuity equation and is therefore conserved throughout the
motion, and $\Gamma=1+1/n$.
In the static case with spherically symmetry, with the space-time
metric of the form, \begin{equation} {\rm d}s^2=e^\nu c^2 {\rm
d}t^2-r^2({\rm d}\vartheta^2+sin^2\vartheta {\rm
d}\varphi^2)-e^\lambda {\rm d}r^2, \end{equation}
the hydrostatic equilibrium condition is derived to be~\citep{OV}
\begin{equation} \frac{1-2GM(r)/c^2r}{P+\rho c^2}r^2 \frac{d
P}{dr}+\frac{GM(r)}{c^2}+\frac{4\pi G}{c^4}r^3P=0,\end{equation} and
\begin{equation} M=\int^R_0 \rho/c^2 \cdot 4\pi r^2 {\rm d}r. \end{equation}

Similar to the Lane-Emden equation of normal stars, re-scale density
$\rho_g$ and radius $r$, as well as $M(r)$, by
\begin{equation} \rho_g=\rho_{gc}\theta^n, ~~~~~r=\xi/A, \end{equation}
\begin{equation}
M(r)=\frac{4\pi \rho_{gc}}{A^3}\upsilon(\xi),\end{equation}
where $$A^2=\frac{4\pi G \rho_{gc}}{(n+1)\alpha c^2},~~~~~
\alpha=\frac{K{\rho_{gc}}^{\frac{1}{n}}}{c^2},$$
and $\rho_{gc}$ is the rest mass density at the center, we can then
obtain
\begin{equation} \frac{1-2(n+1)\alpha \upsilon /\xi}{1+(n+1)\alpha
\theta}\xi^2\theta^\prime +\upsilon +\alpha
\xi^3\theta^{n+1}=0,\end{equation} \begin{equation} \upsilon^\prime
=\xi^2\theta^n(1+n\alpha\theta),\end{equation}
with the initial conditions, \begin{equation}
\theta(0)=1,~~~~~\upsilon(0)=0.\end{equation}

The mass and radius of a star are evaluated at the point when the
density reaches the surface density. In general, the surface of a
star is defined by the position where the pressure is zero, or only
the radial pressure is zero~\citep{hs1997,h2008} in the anisotropic
case where the radial pressure is different from the tangential one.
In fact, the surface in our calculations may not be the real
physical surface, since the pressure could not be zero there.
However, this does not significantly affect the mass and radius we
obtain, because a quark star has a sharp edge, and the pressure and
density decrease to zero in a layer with thickness of few
femto-meters.

\subsection{Quark Stars with QCD Vacuum Energy}

The energy-momentum tensor has the form \begin{equation}
\mathcal{T}_{\mu \nu}=\mathcal{T}_{\mu
\nu}^{\rm{particles}}+\mathcal{T}_{\mu \nu}^{\rm vacuum}.
\end{equation} For a constant vacuum energy, the energy-momentum tensor can be
written as \begin{equation} \mathcal{T}_{\mu
\nu}^{\rm{vacuum}}=\Lambda g_{\mu \nu}, \end{equation} and one finds
\begin{equation} P^{\rm{vacuum}}=\mathcal{T}_{ii}^{\rm
{vacuum}}=-\Lambda, \end{equation} \begin{equation}
\rho^{\rm{vacuum}}=\mathcal{T}_{00}^{\rm{vacuum}}=\Lambda.
\end{equation}

We can infer that no matter what forms the equations of state for
the particles are, the contribution of vacuum is of the above form.
Consequently, we write the equation of state as \begin{equation} P=K
\rho_g^{1+\frac{1}{n}}-\Lambda, \end{equation} \begin{equation}
\rho=\rho_g c^2+n K \rho_g^{1+\frac{1}{n}}+\Lambda.
\end{equation} In this case, the density at surface (where pressure
is zero) should also be non-zero.
It is worth noting that these general equations of state of Eq.(15)
and Eq.(16) could be simplified into a few special ones: the form
without QCD vacuum energy if $\Lambda=0$, the linear (relativistic)
form if $\rho_g=0$, and the non-relativistic form if the second term
in Eq.(16) is neglected (see the next sub-section for more
discussions).

The hydrostatic equilibrium conditions are also determined by Eq.(4)
and Eq.(5).
But in this case, we cannot derive differential equations for
density and radius, and can only numerically calculate the structure
of a quark star from the center to the surface to obtain the mass
and radius.

\subsection{Comparison with MIT bag model}

In the MIT bag model, quark matter are composed of massless up and
down quarks, massive strange quarks, and electrons. Quarks are
combined together by an extra pressure, denoted by the bag constant
$B$, which is the vacuum energy density similar to the
$\Lambda$-parameter in the polytropic model. For the comparison, we
apply the formulae given by Alcock~\citep{Alcock1986} to calculate
the equation of state, with strange quark mass $m_s=250 \rm MeV$ and
the strong coupling constant $\alpha_s=0.6$ for indications.

Note that the $K$-parameter in \S2.2 could be calculated in the way
of, from Eq.(15),
\begin{equation} K=\Lambda \cdot \rho_{\rm sur}^{-(1+\frac{1}{n})},
\end{equation} where $\rho_{\rm sur}$ is the surface density, since at the
surface the pressure is zero.
This density should be determined by the behavior of the elementary
strong interaction. In the calculation, we assume that $\Lambda$ has
the same value as the bag constant $B$ of MIT bag model, and choose
the surface density, $\rho_{\rm sur}=1.5\rho_0$, where $\rho_0$ is
the nuclear matter density.
The $K$-parameter could be {\em larger} if the quark
self-confinement effect is included, and the value determined by
Eq.(17) is the minimal and the maximum masses presented in Fig. 2
and Fig. 3 could also be higher.
For the sake of simplicity, we suppose a value of $K$ from Eq.(17),
and assume the $K$-parameter is the same for both cases with and
without vacuum energy under a same polytropic index $n$.
If quark-cluster inside a quark star are very massive, the kinetic
energy density could be negligible compared to the rest mass energy
density, i.e., the equation of state could be non-relativistic, and
the total energy density would include only the rest mass energy
density\footnote{
An example similar to this non-relativistic equation of state is of
the matter in white dwarfs where the energy density is dominated by
the rest mass of nuclei, while the electron gas contributes to the
pressure.
} in Eq.(16).
%
\begin{figure}
\begin{center}

  \includegraphics[width=6.5cm]{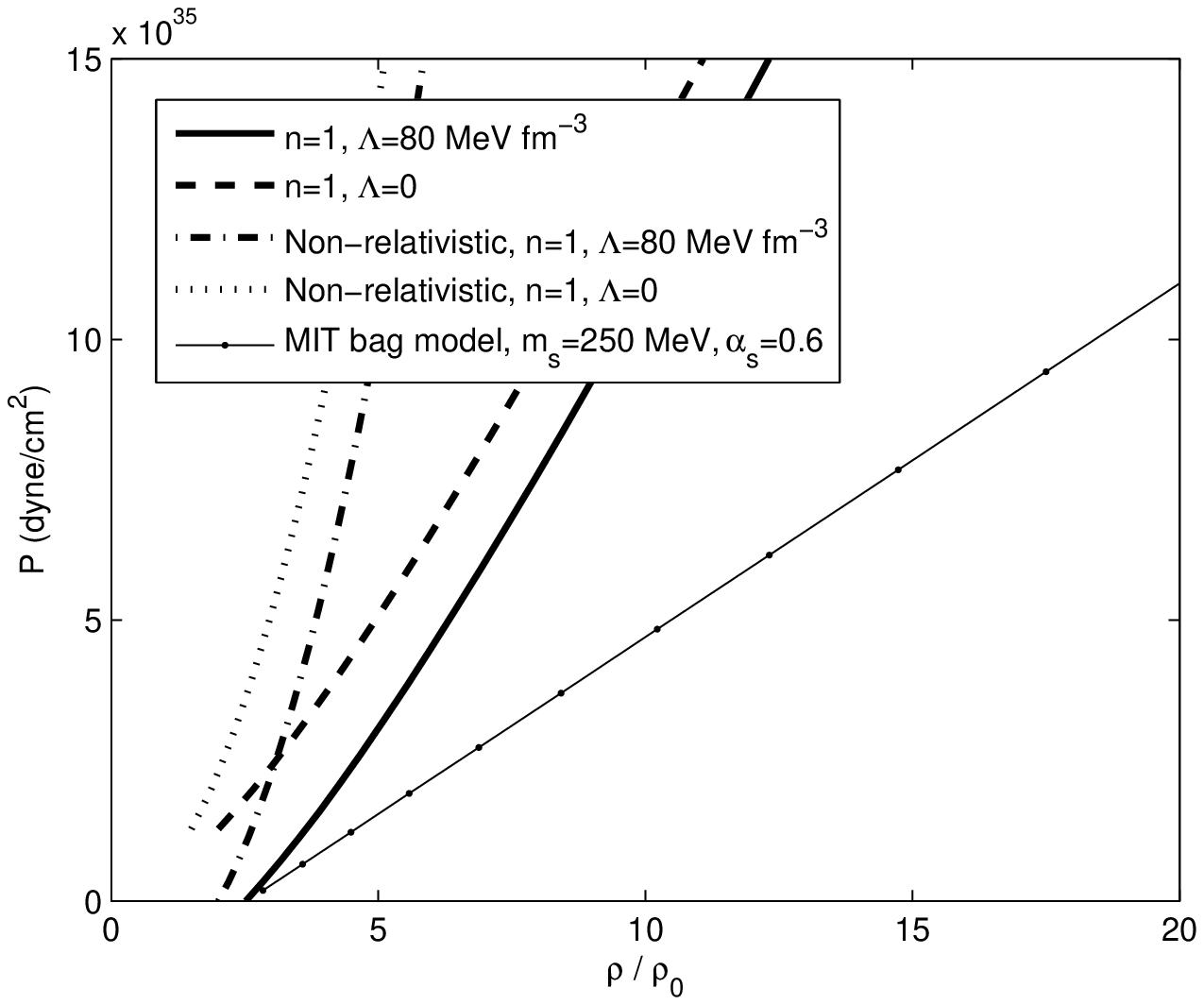}
  \includegraphics[width=6.5cm]{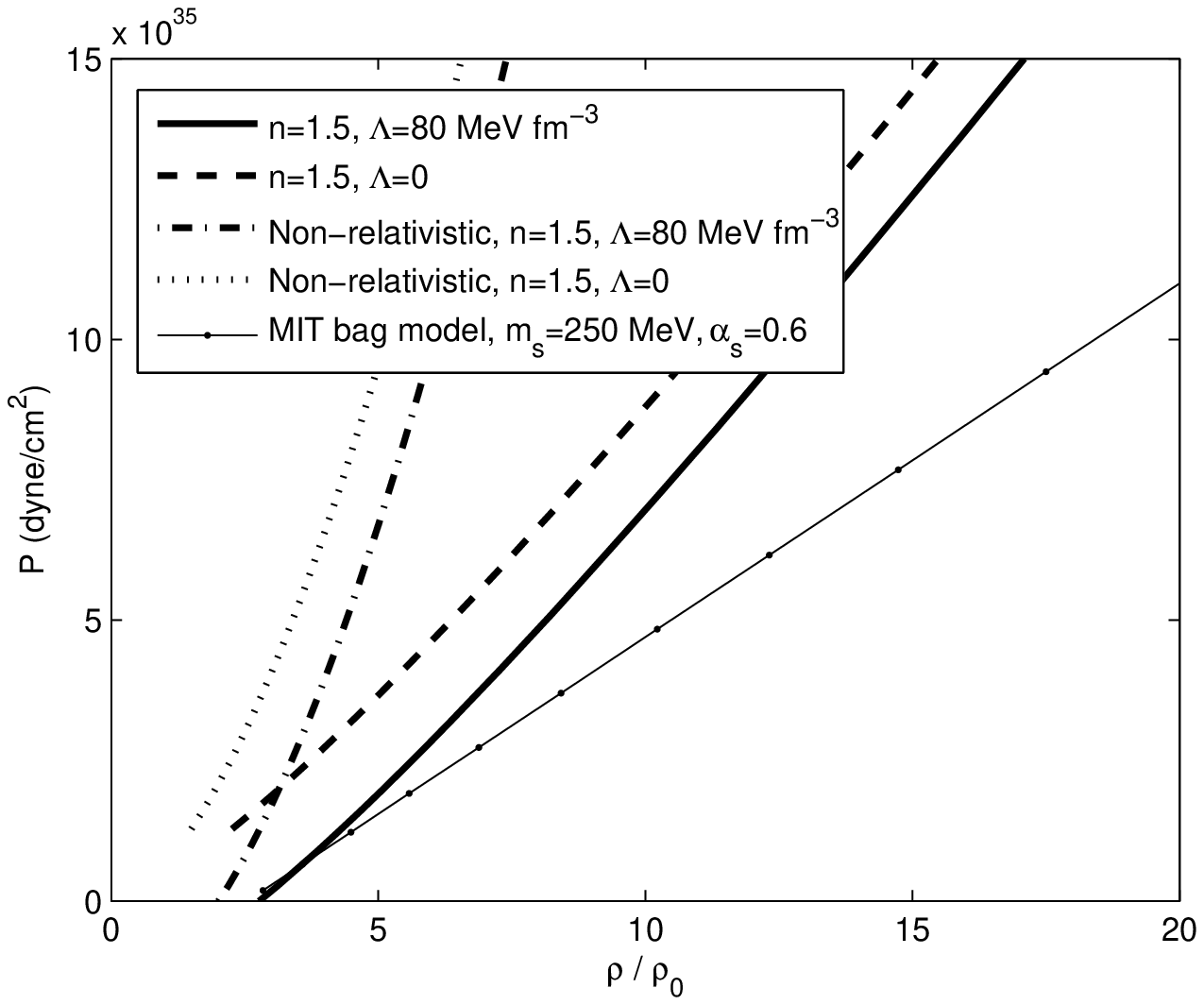}
  \includegraphics[width=6.5cm]{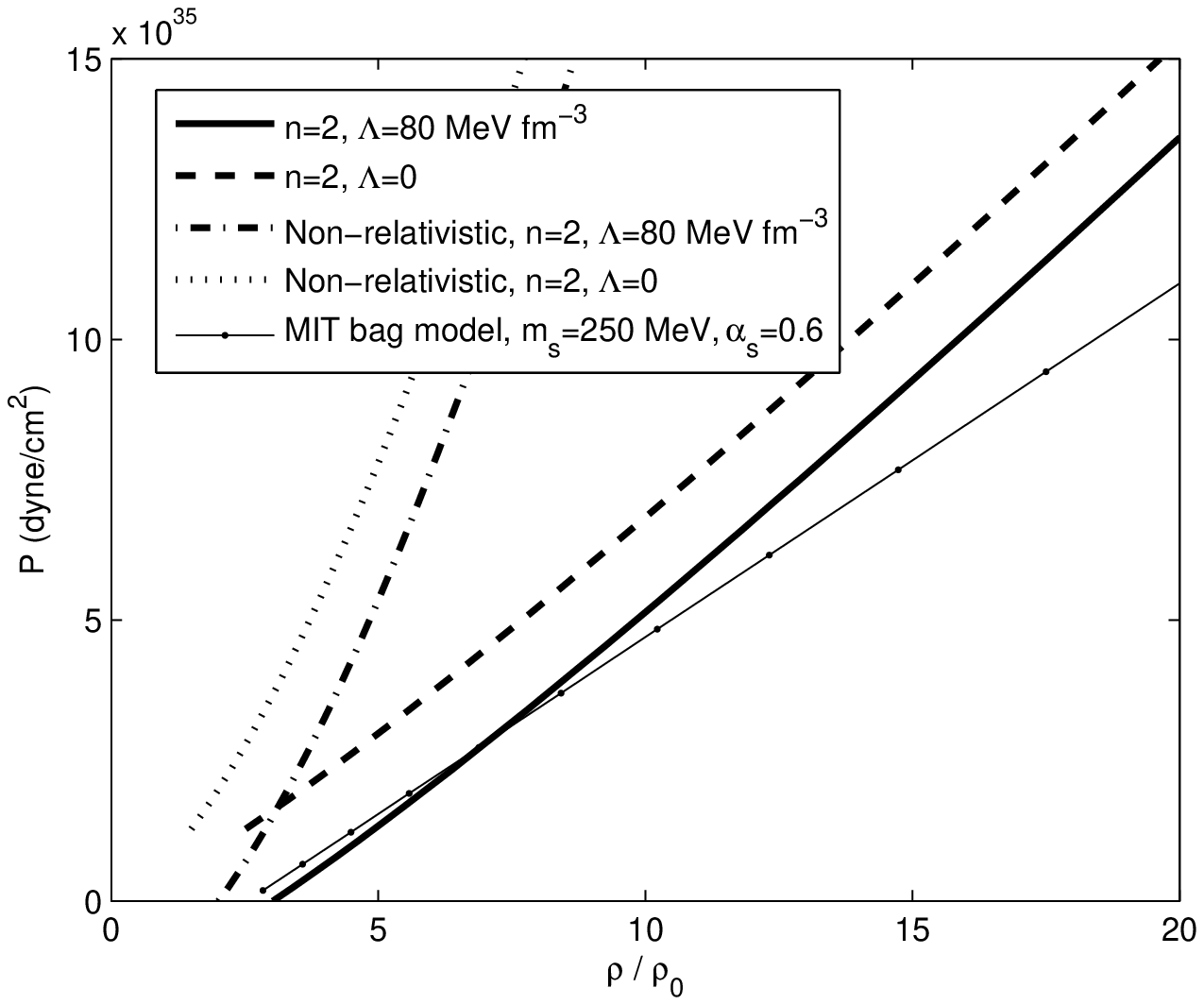}
  \includegraphics[width=6.5cm]{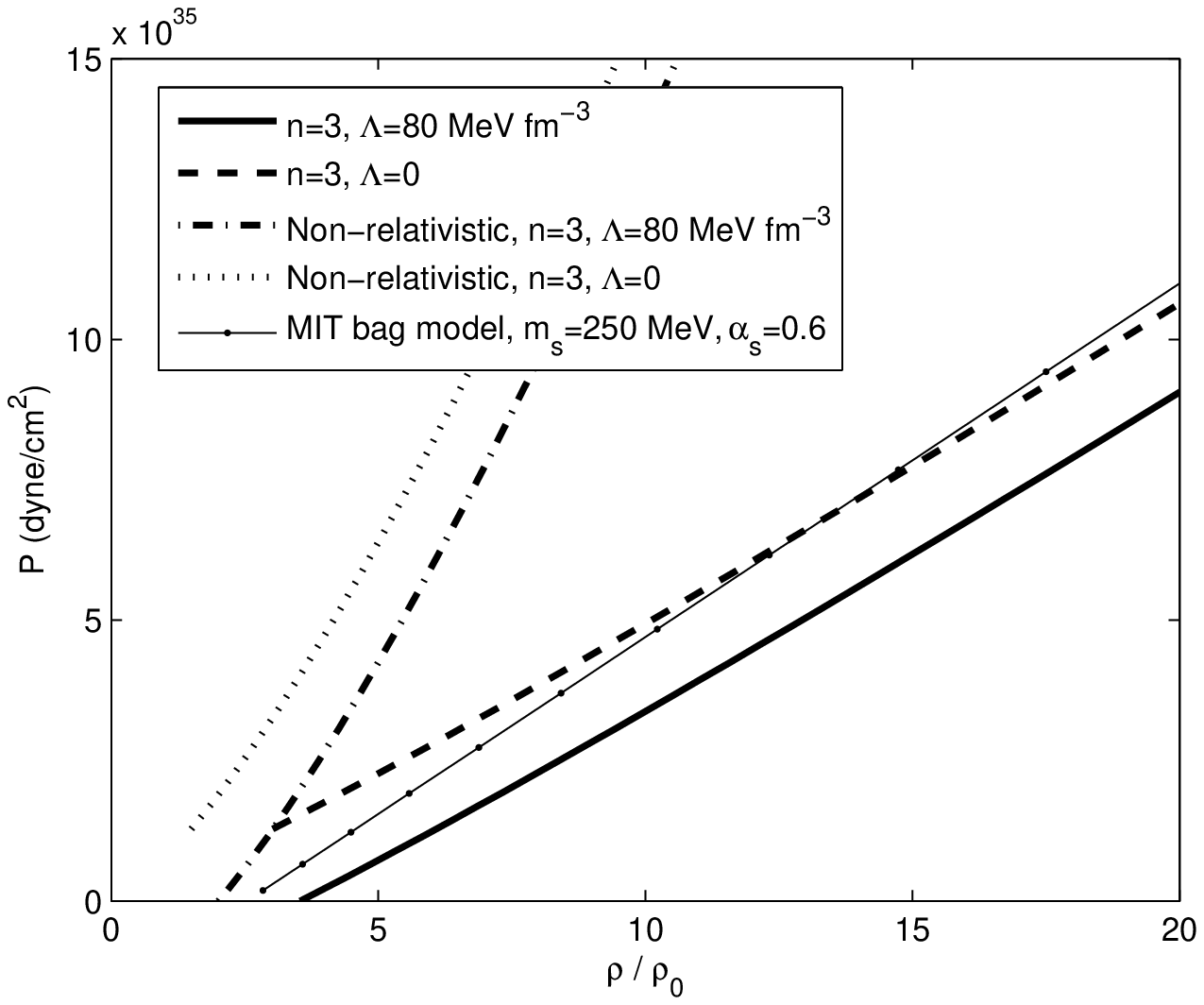}

\end{center}
\caption{To compare the equations of state discussed, including the
polytropic states with $\Lambda=80 \rm MeV$ (solid lines),
$\Lambda=0$ (dashed lines), the corresponding non-relativistic case
with $\Lambda=80 \rm MeV$ (dash-dotted lines), $\Lambda=0$ (dotted
lines), and that derived in the MIT bag model with the mass of
strange quark $m_s=250$ MeV and the strong coupling constant
$\alpha_s=0.6$ (thin lines with dots), for a given surface density
$\rho_{\rm sur}=1.5\rho_0$. Here and in the following figures,
$\rho_0$ is the nuclear saturation density. It is evident that
polytropic equations of state are stiff.} \label{figure 1}
\end{figure}

The equations of state discussed above are shown in Fig.1.
An obvious conclusion is that the equations of state of polytropic
form could be stiffer than that of the MIT bag model, especially in
the non-relativistic case.
A stiffer equation of state would lead to a larger maximum mass of
quark stars.

\subsection{Gravitational energy in general relativity}

The gravitational energy in general relativity was calculated
by~\cite{Tooper}. In general relativity, the integrating of space
volume should be different from that in the Newtonian gravity due to
the space-time curvature around a massive star.
The proper energy, $E_0$, is obtained by integrating the energy
density over elements of proper spatial volume,
\begin{equation} E_0=E_{0g}+E_{0k},\end{equation}
where the rest energy, $E_{0g}$, of the system and its microscopic
kinetic energy, $E_{0k}$, are
\begin{equation} E_{0g}=M_{0g}c^2=4\pi\int^R_0 \rho_g c^2 e^{\lambda/2}r^2{\rm
d}r, \end{equation} \begin{equation} E_{0k}=4\pi\int^R_0
nPe^{\lambda/2}r^2{\rm d}r, \end{equation} where $\lambda$ can be
calculated by
\begin{equation} e^{-\lambda}=1-\frac{2GM}{c^2r}. \end{equation}
The total energy of a star with mass $M=M(R)$ is $E=Mc^2$, and its
gravitational energy, $\Omega$, is the difference between the total
energy and the proper energy,
\begin{equation} \Omega=Mc^2-E_0.\end{equation}

\section{Stars with an anisotropic pressure}

Fluid within inhomogeneous pressure is imperfect, and we will
consider only the case of spherical symmetry, that the tangential
and radial pressure are not equal.
In this case the hydrostatic equilibrium condition
reads~\citep[e.g.,][]{xty06}
\begin{equation} \frac{1-2GM(r)/c^2r}{P+\rho c^2}(r^2 \frac{d
P}{dr}-2\varepsilon r p)+\frac{GM(r)}{c^2}+\frac{4\pi
G}{c^4}r^3P=0,\end{equation}
where $\varepsilon$ is defined by $P_\perp=(1+\varepsilon)P$, and
$P$ is the radial pressure and $P_\perp$ is the tangential one.

Combine the hydrostatic equilibrium condition and equation of state,
one can calculate the structures of quark stars with and without QCD
vacuum energy.

\section{Numerical results}

Based on the formulae presented in \S2.1, the mass-radius relations
for various index, $n$, can be calculated. We are applying the
Runge-Kutta method of order 4 to solve the differential equations,
until the density reaches the surface density. For the case in
\S2.2, we numerically calculate from the center to the surface and
obtain then the mass and radius.

It is worth noting that the non-zero surface density of quark stars
play an important role in the computation.
This density should be determined by the behavior of the elementary
strong interaction, and is then an uncertain parameter. In the
calculation as following, we choose the surface density, $\rho_{\rm
sur}=1.5\rho_0$, where $\rho_0$ is the nuclear matter density.

In calculating the gravitational energy, we numerically integrate
from the center to the surface for both cases with and without QCD
vacuum energy.

\subsection{Mass-radius relations for stars of perfect fluid}

The mass-radius curves for both cases with and without QCD vacuum
energy are shown in Fig.2.
%
\begin{figure}
\begin{center}

  \includegraphics[width=6.5cm]{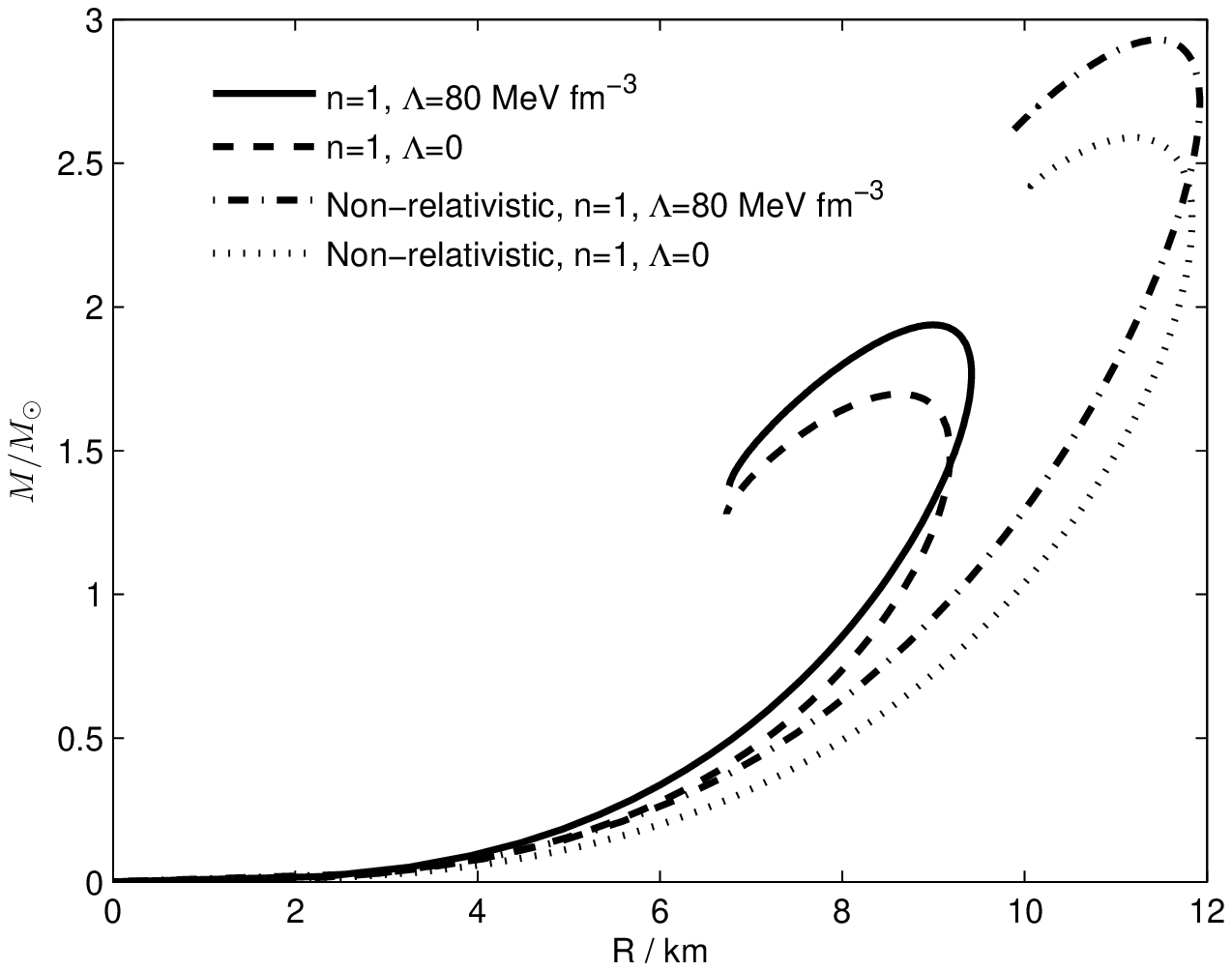}
  \includegraphics[width=6.5cm]{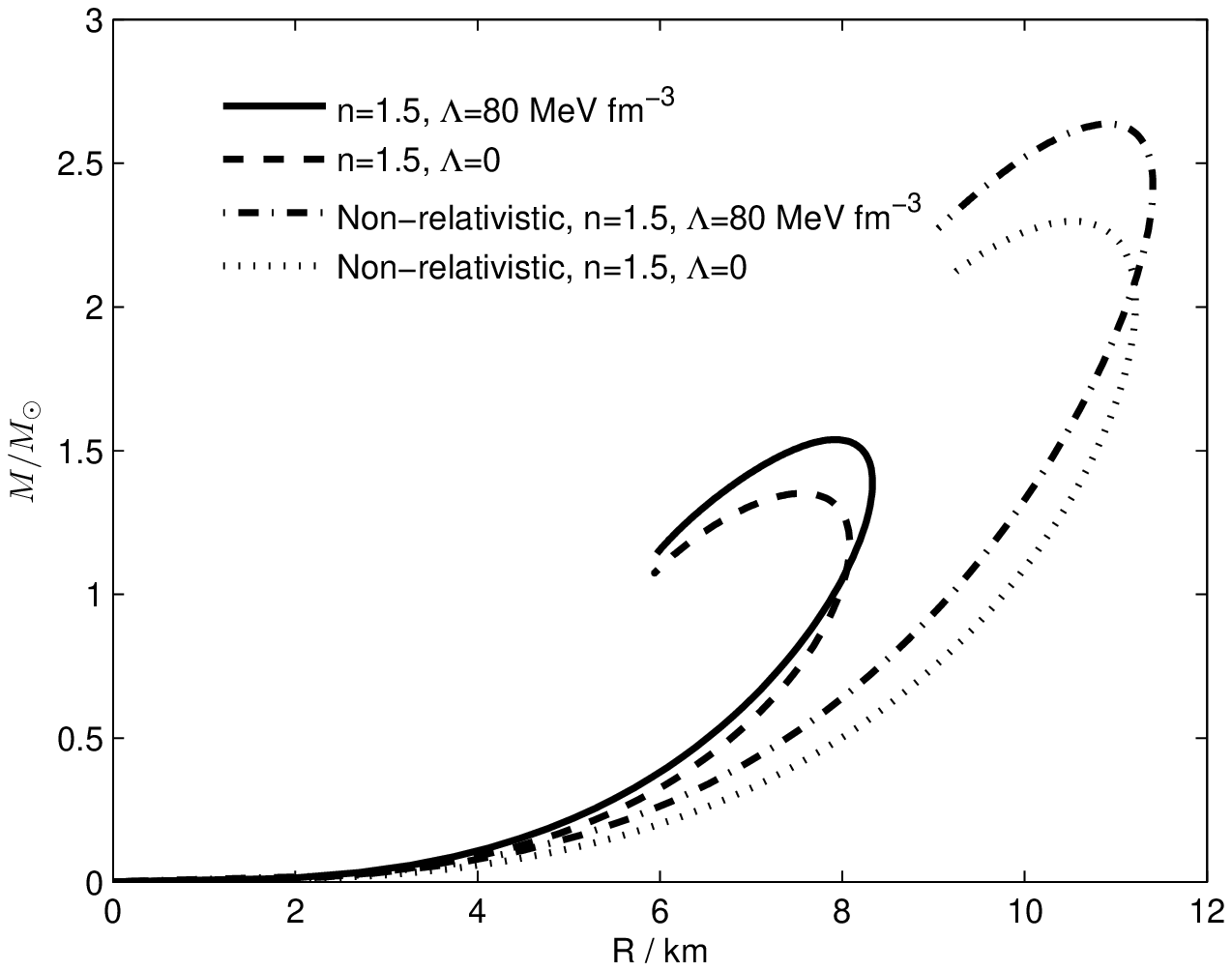}
  \includegraphics[width=6.5cm]{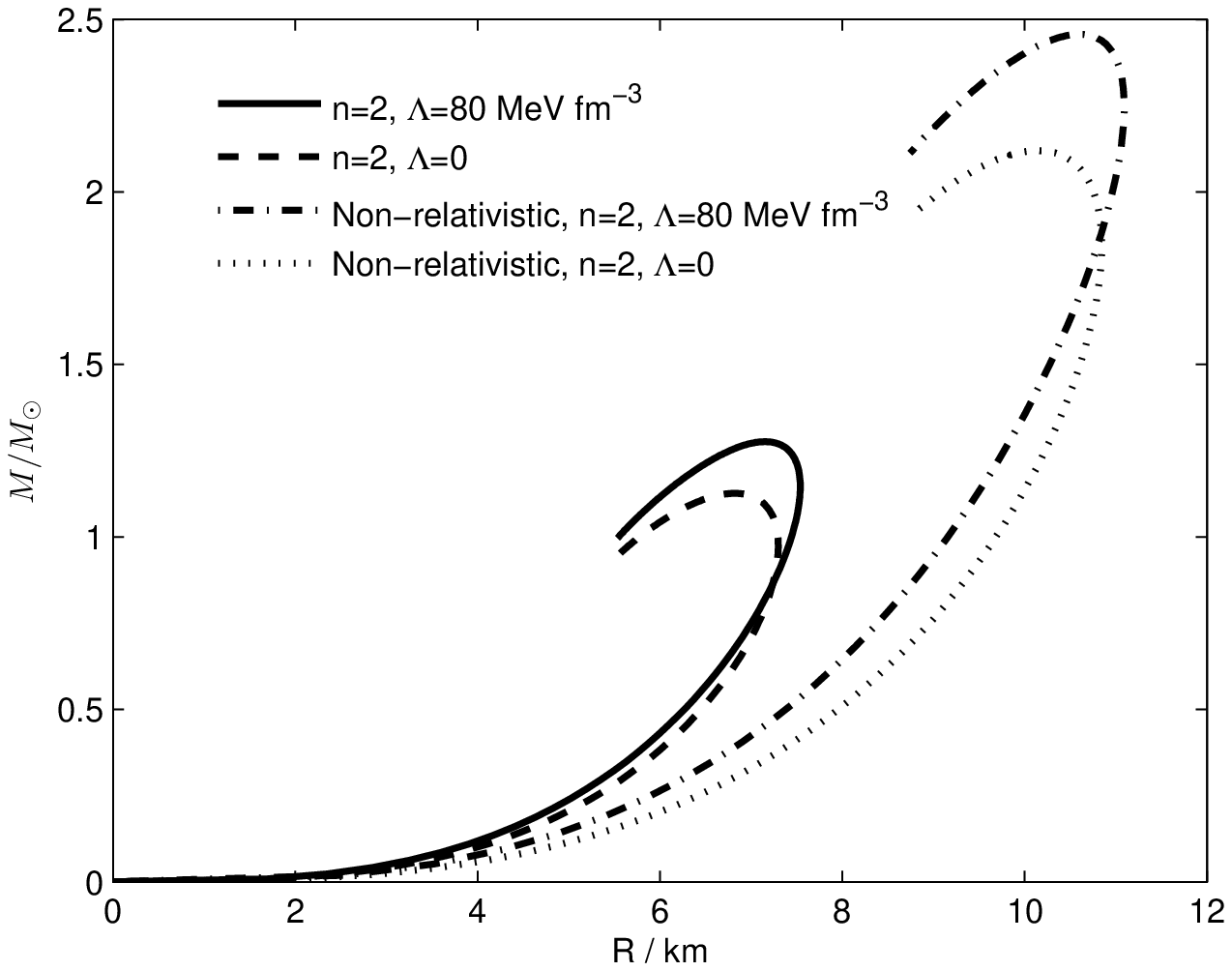}
  \includegraphics[width=6.5cm]{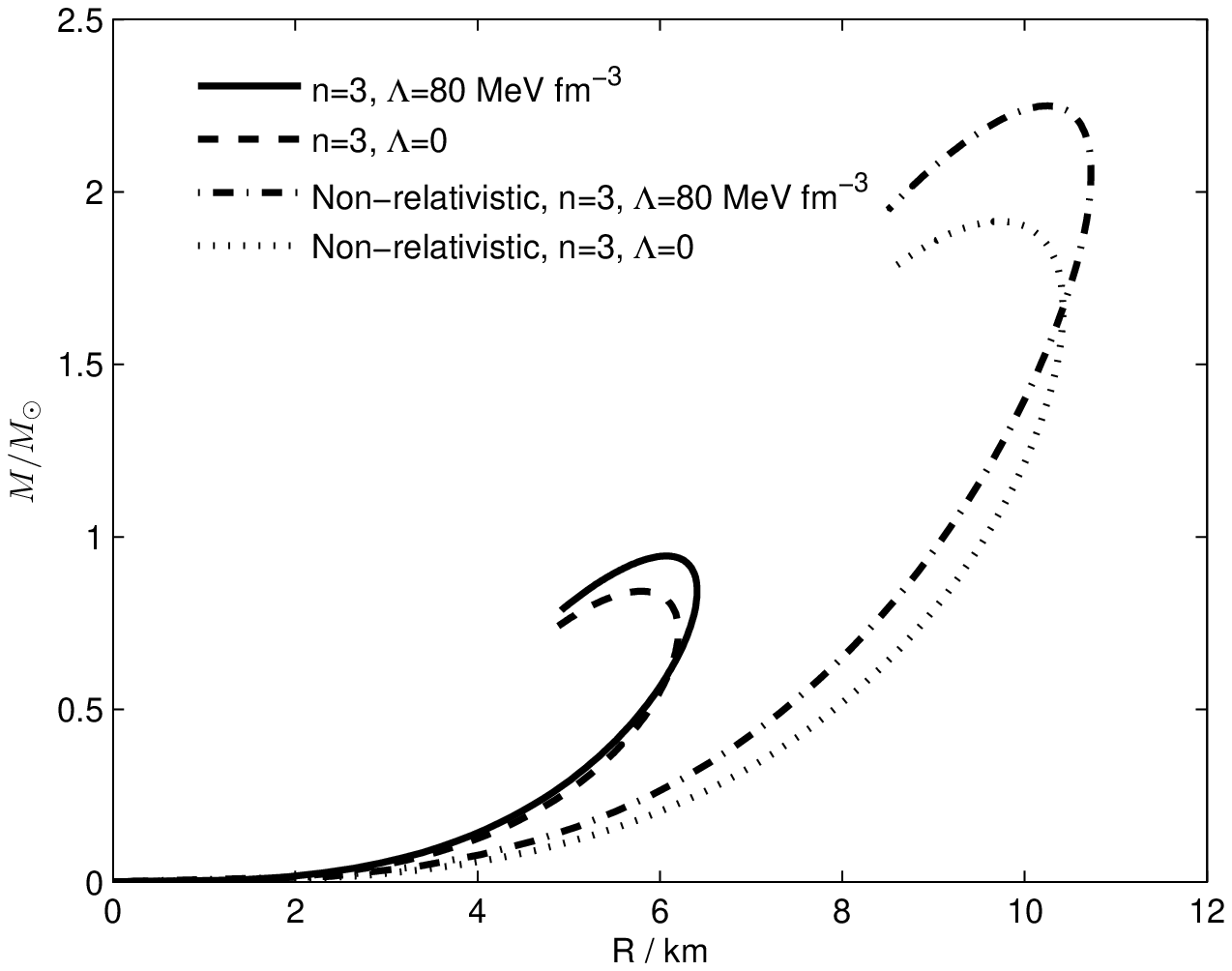}

\end{center}
\caption{Mass-radius relations for different polytropic indices,
$n$, with $\rho_{\rm sur}=1.5\rho_0$. Solid lines are
$\Lambda=80\rm{MeV fm^{-3}}$, dashed lines are for $\Lambda=0$,
dash-dotted lines are for non-relativistic case with
$\Lambda=80\rm{MeV fm^{-3}}$, and dotted lines are for
non-relativistic case with $\Lambda=0$.} \label{figure 2}
\end{figure}
%
It is evident from the calculation that the maximum mass of quark
star decreases as the index, $n$, increases.
This is understandable. A small $n$ means a large $\Gamma$, and the
pressure is relatively lower for higher values of $n$. Lower
pressure should certainly support a lower mass of star.

It could have observational implications that the maximum mass of
quark stars with polytropic equations of state are larger than that
derived in conventional model.
Recently the results of 19 years of Arecibo timing for two pulsars
in the globular cluster NGC 5904 (M5) had been reported by Freire et
al.~\citep{Freire}. They confirmed that for one of the binary
pulsars (M5B) the mass is $2.08\pm0.19 M_\odot$ in $1\sigma$, and
concluded that this mass for the pulsar would exclude most ``soft''
equations of state for dense neutron matter.
However, a quark star in the polytropic model could be more massive
than in previously derived realistic models (e.g., the MIT bag
model) because of a stiffer equation of state, and the maximum mass
could be larger than $2M_\odot$.
It is worth emphasizing that the maximum mass of quark star depends
on the value of $K$-parameter. Although in this paper we derive its
minimum value under some assumptions, its real value is still
uncertain. On this point of view, even the pulsars of masses larger
than $1.4M_\odot$ have been observed, the case of $n=3$ could not be
ruled out because the value of $K$ could be larger than the value we
use.

{\em On the gravitational stability.}
A polytropic star, with a state equation of $P\propto \rho^\Gamma$,
supports itself against gravity by pressure, $PR^2$ (note: the
stellar gravity $\propto M/R^2\propto\rho R$).
Certainly, a high pressure (and thus large $\Gamma$ or small $n$) is
necessary for a gravitationally stable star, otherwise a star could
be unstable due to strong gravity.
Actually, in the Newtonian gravity, a polytropic normal star (with
$\rho_{\rm sur}=0$) is gravitationally unstable if $n>3$, but the
star should be still unstable if $n=3$ when the GR effect is
included~\citep{Shapiro Teukolsky}.

A polytropic quark star with non-zero surface density or with QCD
vacuum energy, however, can still be gravitationally stable even if
$n\geq 3$.
A quark star with much low mass could be self-bound dominantly, and
the gravity is negligible (thus not being gravitationally unstable).
As the stellar mass increases, the gravitational effect becomes more
and more significant, and finally the star could be gravitationally
unstable when the mass increases beyond the maximum mass. In order
to see the central density-dependence of stability, the calculated
mass-central density curves are shown in Fig. 3.
%
\begin{figure}
\begin{center}

  \includegraphics[width=6.5cm]{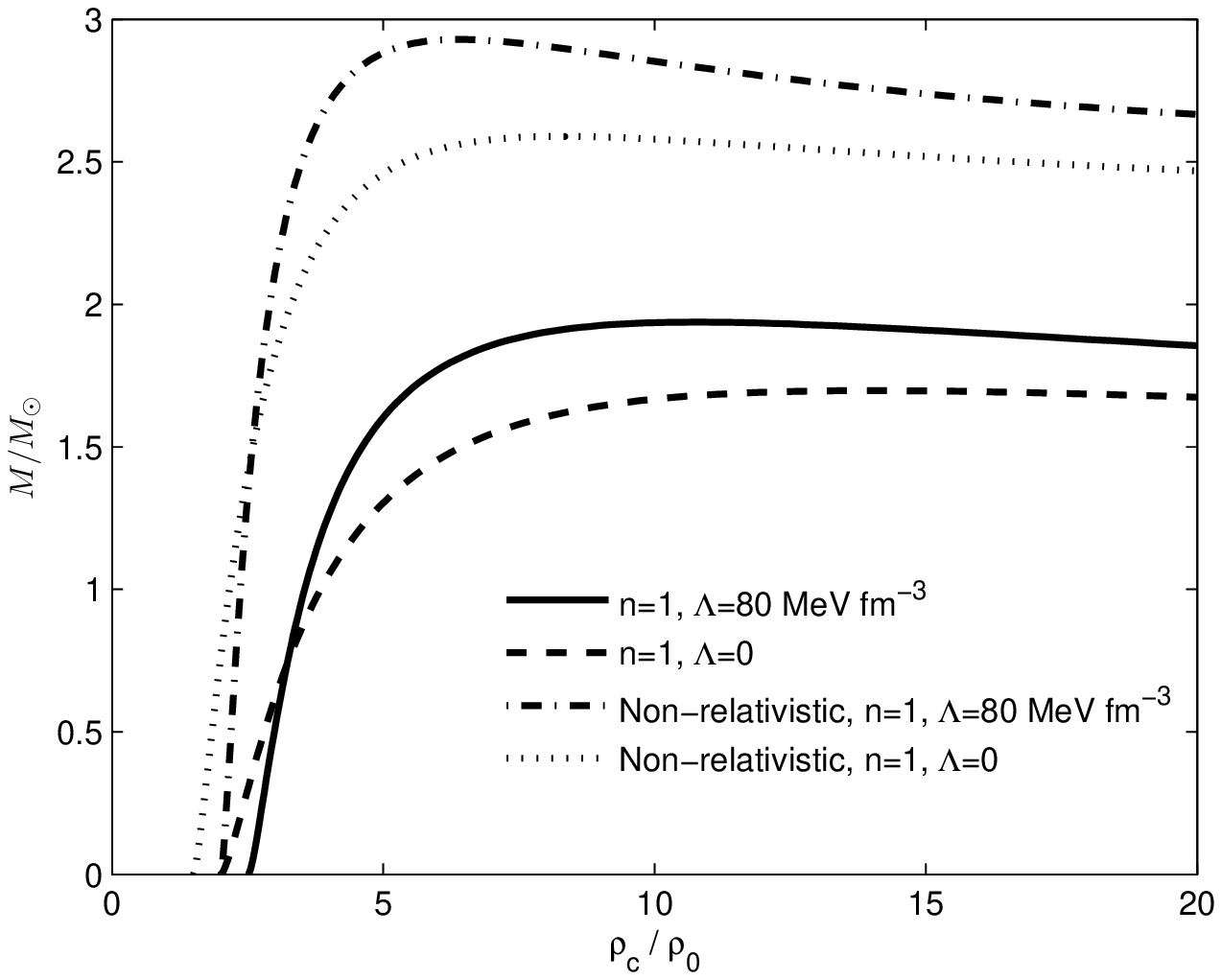}
  \includegraphics[width=6.5cm]{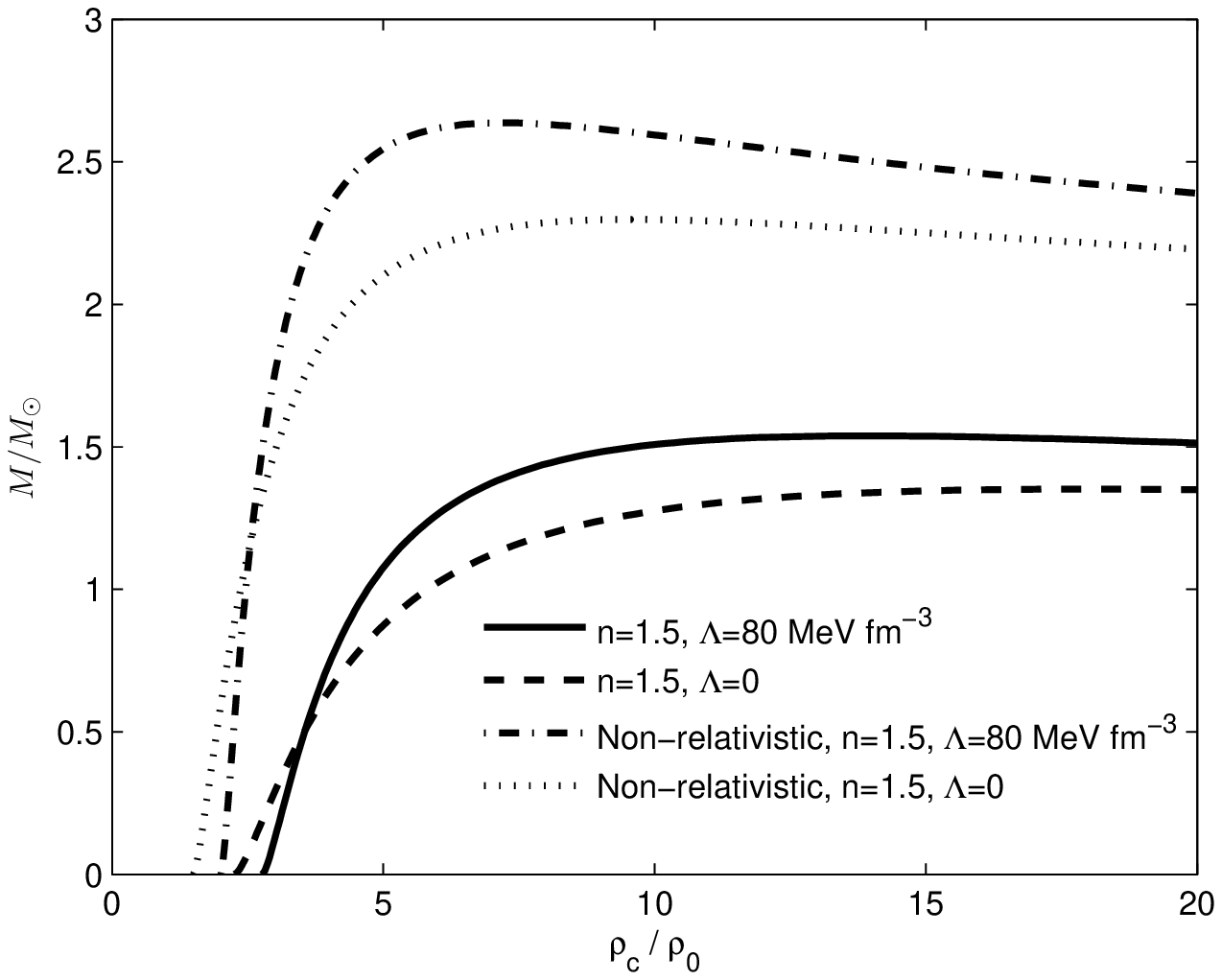}
  \includegraphics[width=6.5cm]{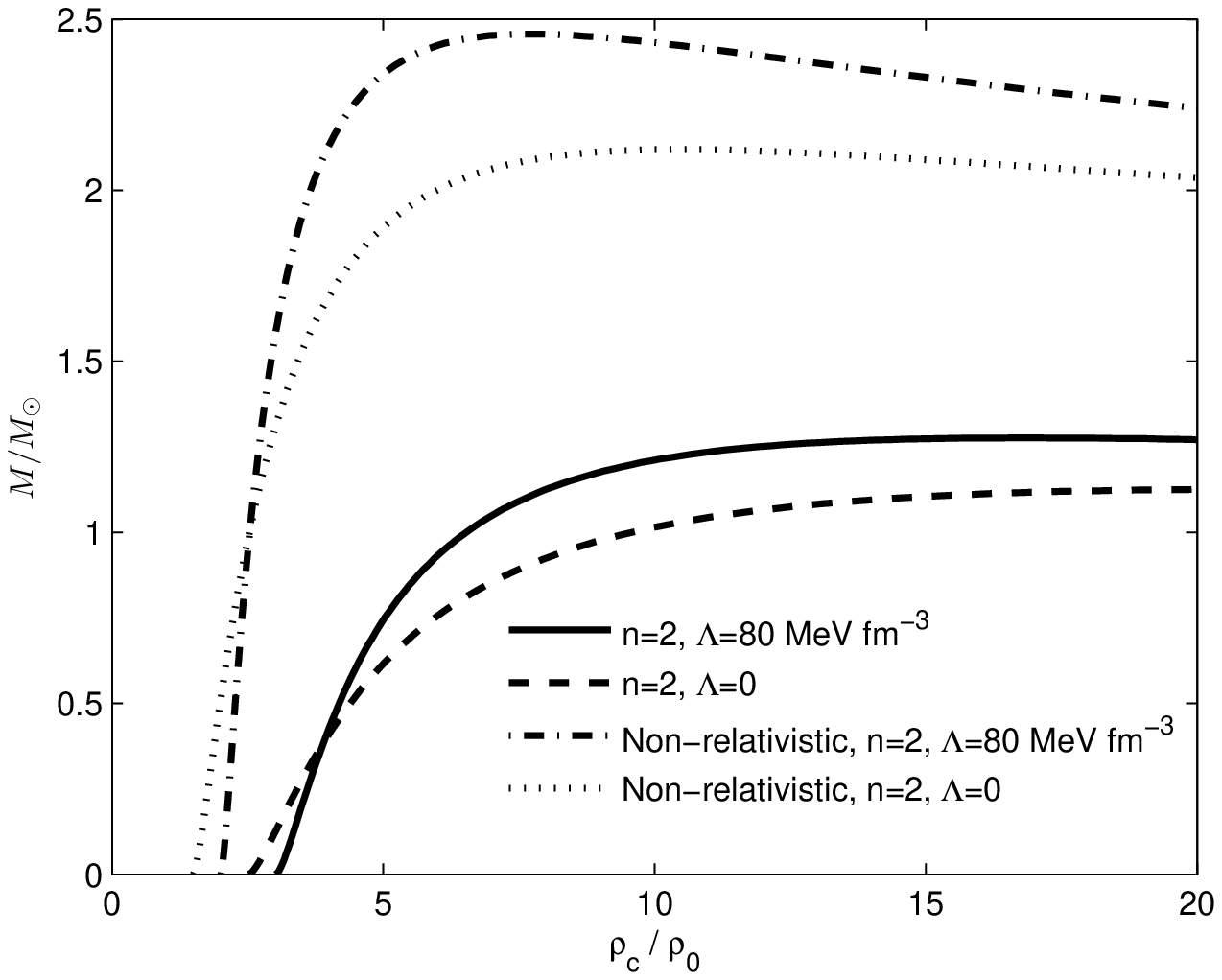}
  \includegraphics[width=6.5cm]{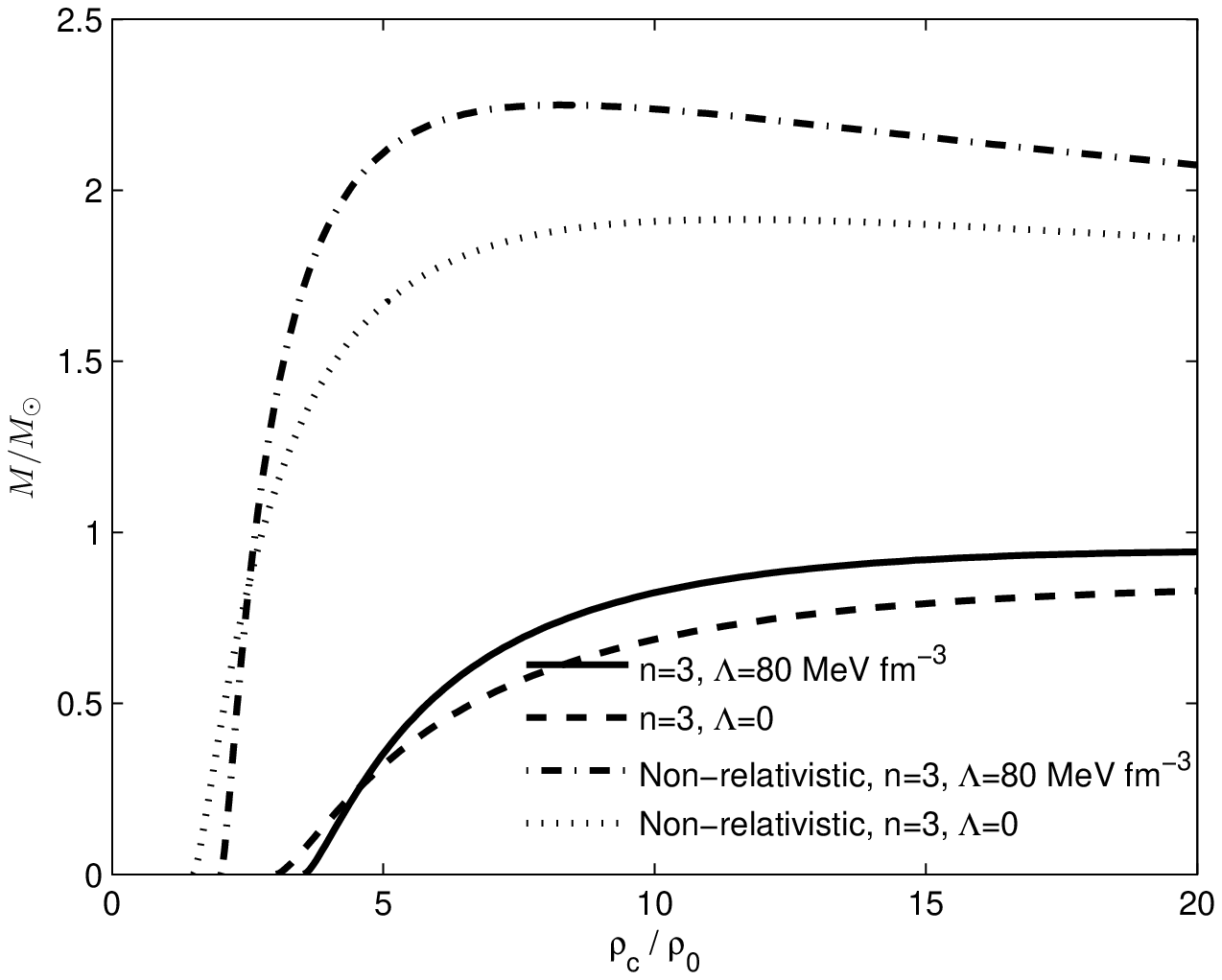}

\end{center}
\caption{Mass-central density relations for different polytropic
indices, $n$, with $\rho_{\rm sur}=1.5\rho_0$. Solid lines are
$\Lambda=80\rm{MeV fm^{-3}}$, dashed lines are for $\Lambda=0$,
dash-dotted lines are for non-relativistic case with
$\Lambda=80\rm{MeV fm^{-3}}$, and dotted lines are for
non-relativistic case with $\Lambda=0$. Stars with central densities
greater than that of stars with maximum masses are gravitationally
unstable.} \label{figure 3}
\end{figure}
%

{\em Sound speed.}
From the $P-\rho$ relation in \S2.1, we can derive the ratio of
sound speed to speed of light, which cannot be greater than 1,
\begin{equation}
(\frac{v_s}{c})^2=\frac{{\rm d}P}{{\rm
d}\rho}=\frac{n+1}{n}\frac{P}{\rho +P}\leq 1.
\end{equation} Similarly, also from the relation given in \S2.2 we can come to \begin{equation}
(\frac{v_s}{c})^2=\frac{{\rm d}P}{{\rm
d}\rho}=\frac{n+1}{n}\frac{P+\Lambda}{\rho +P}\leq 1.
\end{equation} Both of the two inequations lead to
 \begin{equation} (1-n^2)K \rho_g^{\frac{1}{n}} \leq n c^2. \end{equation}
The equation above holds if $n \geq 1$, that means that the
causality keeps for $n \geq 1$.

\subsection{Gravitational energy released during a star quake}

Based on various manifestations of pulsar-like stars, a solid state
of cold quark matter was conjectured~\citep{xu03}.
A solid stellar object would inevitably result in starquakes when
strain energy develops to a critical value, and a huge of
gravitational and elastic energies would then be released.
One way to accumulate both shear and bulk forces in a solid quark
star is during an accretion process: strain develops remarkably in
massive stars, for which the gravitational effect is not negligible.
A solid star could additionally support the accreted matter against
gravity by these forces, unless the forces become so strong that a
star-quake occurs.
This is the so-called {\em AIQ} (Accretion-Induced star-Quake)
mechanism proposed previously~\citep{xty06,xu07}, which might be
responsible to the bursts (even the supergiant flares) and glitches
observed in soft $\gamma$-ray repeaters/anomalous X-ray
pulsars~\citep[see a recent review by][]{Mereghetti08}.

How to calculate the energy released during an AIQ?
Theoretically, anisotropic fluid stars could be introduced for,
e.g., the presence of type 3A superfluid~\citep{Kippenhahn Weigert},
different kind of phase transitions~\citep{Sokolov}, and pion
condensation~\citep{Sawyer}.
For the general relativistic configurations, when the interactions
between particles could be treated relativistically, the fluid could
also be anisotropic~\citep{Ruderman}.
Previous theoretical results for anisotropic fluid, in a simple case
with spherical symmetry, could still be adaptable to estimate the
AIQ-released energy of a solid star.

Based on the analytical study of anisotropic matter, Harko and
Mak~\citep{Harko 2,Harko 1} discussed the constraints for the
anisotropic parameter defined in \S3, $\varepsilon$, and presented
an exact analytical solution for the gravitational equations of a
static spherically symmetric anisotropic quark matter
star~\citep{Harko 3}. It is found that the $\varepsilon$-value could
be as high as $10^{-2}$.
Applying the formulae in \S2.4 and \S3, we can obtain the
gravitational energy released during a star-quakes in both case with
and without QCD vacuum energy, for example, with $n=1$. The
gravitational energy released for quark stars of linear equation of
state with two different vacuum energy has been calculated in
~\cite{xty06}.
We approximate that the rest energy (corresponds to the total baryon
mass) $E_{0g}$ does not change during a star quake, $E_{0g}=M_0
c^2$, since the released energy is much smaller than $E_{0g}$.

The gravitational energy difference between stars with $\varepsilon
\neq 0$ and with $\varepsilon=0$ are shown in Fig. 4.
Three supergiant flares from soft $\gamma$-ray repeaters have been
observed, with released photon energy being order of $\sim 10^{47}$
ergs.
Our numerical results imply that for all the parameters we chosen,
the released energy could be as high as the observed.
%

\begin{figure}
\begin{center}

  \includegraphics[width=8cm]{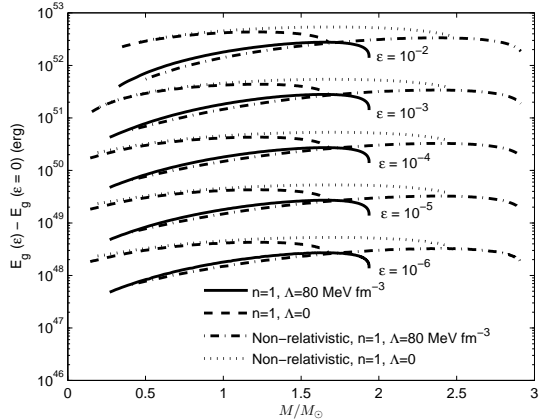}

\end{center}
\caption{The gravitational energy difference between stars with and
without anisotropic pressures, which may be released during
sequential star quakes. Solid lines are $\Lambda=80\rm{MeV
fm^{-3}}$, dashed lines are for $\Lambda=0$, dash-dotted lines are
for non-relativistic case with $\Lambda=80\rm{MeV fm^{-3}}$, and
dotted lines are for non-relativistic case with
$\Lambda=0$.}\label{figure 4}
\end{figure}

\section{Conclusions and Discussions}

Because of the difficulty to obtain a realistic state equation of
cold quark matter at a few nuclear densities, we have tried to apply
polytropic equations of state to model quark stars in this paper.
The polytropic equations of state to quark stars are studied in two
separated cases: the vacuum inside and outside quark matter is the
same or not.
In addition, the quark-clustering could lead to the non-relativistic
equation of state.
The differences between the those cases could be significant in the
mass-radius relations, and may be tested by observations. It could
additionally provide a way to probe the properties of QCD vacuum.

The polytropic equations of state are stiffer than that derived in
previous realistic models, so they could lead to more massive quark
stars with masses $>2M_\odot$. Consequently, even when some massive
pulsars have been observed, it still can not rule out the
possibility that pulsar-like stars are quark stars.
Though a normal star with zero surface density can only be
gravitationally stable if the polytropic index $n<3$, a quark star
with non-zero surface density could still be stable even if $n\geq
3$.
A solid quark star may break if its strain energy develops to a
critical value, and we calculate the gravitational energy released
during quakes and find that the energy could be as high as $\sim
10^{47}$ ergs if the anisotropic parameter, $\varepsilon$, could be
order of $10^{-6}$. Such a huge of energy would be liberated during
an AIQ (accretion-induced star-quake) process, to be probably
responsible to the bursts and glitches observed in soft $\gamma$-ray
repeaters/anomalous X-ray pulsars.
The general relativity effect has been included to simulate the
polytropic quark stars.

The nature of pulsars is unfortunately still a matter of
controversy, even more than 40 years after the discovery of pulsar.
Although quark stars cannot be ruled out, both theoretically and
observationally, and pulsars are potential idea laboratories to
study the elementary strong interaction, we are lacking a {\em
general} framework in which theoretical stellar models could be
tested by observations. Polytropic quark star model is the one we
try to establish.
Future advanced observations may help to constrain the uncertain
parameters, e.g., the polytropic index $n$, the coefficient $K$, the
surface density $\rho_{\rm sur}$, and even the vacuum energy
$\Lambda$.

One of the daunting challenges nowadays is to understand the
fundamental strong interaction between quarks, especially the QCD in
the low-energy limit, since the coupling is asymptotically free in
the limit of high-energy.
The state of cold matter at a few nuclear densities is still an
unsolved problem in the low-energy QCD.
In effective QCD models, BCS-type quark pairing was proposed to form
at a Fermi surface of cold quark matter, and the shear moduli of the
rigid crystalline color super-conducting quark matter could be 20 to
1000 times larger than those of neutron star crusts~\citep{mrs07}.
However, quark clusters are phenomenologically suggested to form in
cold quark matter~\citep{xu03}. The state of such cold quark matter
might be approximated by polytropic equations of state since one may
draw naively an analogy between the clusters in quark matter and the
nucleus in normal matter.
Certainly, it would be very interesting to observationally
distinguish between those two kinds of solid quark matter.


{\em Acknowledgments}:
We acknowledge useful discussions at the pulsar group of PKU, and
thank Dr. David Blaschke, Dr. J\"urgen Schaffner-Bielich and an
anonymous referee for their comments and suggestions. This work is
supported by NSFC (10573002, 10778611), the Key Grant Project of
Chinese Ministry of Education (305001), and by LCWR (LHXZ200602).

\end{document}